\documentclass[apj]{emulateapj}
\usepackage{natbib}
\usepackage{hyperref}
\hypersetup{
    colorlinks=true,
    linkcolor=blue,
    citecolor=blue,
    filecolor=blue,
    urlcolor=blue
}
\usepackage{enumitem}

 \providecommand{\adsurl}[1]{\href{#1}{ADS}}
\usepackage{amsmath,amssymb}
\usepackage{mathptmx}
\usepackage{mathtools}
\DeclareMathAlphabet{\pazocal}{OMS}{zplm}{m}{n}

\defcitealias{Bialy2017}{BBS}
\defcitealias{VazquezSemadeni2001}{VG}

\begin{document}

\title{
The driving scale -- density decorrelation scale relation in a turbulent medium
}
 \author{Shmuel Bialy}   
  \affiliation{Harvard-Smithsonian Center for Astrophysics, 60 Garden St., Cambridge, MA, 02138}
  \author{Blakesley Burkhart}
 \affiliation{Department of Physics and Astronomy, Rutgers, The State University of New Jersey, 136 Frelinghuysen Rd, Piscataway, NJ 08854}
 \affiliation{Center for Computational Astrophysics, Flatiron Institute, 162 Fifth Avenue, New York, NY 10010}

\slugcomment{Accepted for publication in ApJ.~Letters}

\begin{abstract}
Density fluctuations produced by supersonic turbulence are of great importance to astrophysical chemical models. A property of these density fluctuations is that the two point correlation function decreases with increasing scale separation.
The relation between the density decorrelation length scale ($L_{\rm dec}$)
and the turbulence driving scale ($L_{\rm drive}$) determines how turbulence affects the density and chemical structures in the interstellar medium (ISM), and is a key component for using observations of atomic and molecular tracers to constrain turbulence properties.
We run a set of numerical simulations of supersonic magnetohydrodynamic turbulence, 
with different sonic Mach numbers ($\mathcal{M}_s=4.5, 7$)
, and driven on varying scales (1/2.5, 1/5, 1/7) the box length.
We derive the $L_{\rm dec}-L_{\rm drive}$ relation as a function of Mach number, driving scale, and the orientation of the line-of-sight (LOS) in respect to the magnetic-field. 
 We find that the mean ratio $L_{\rm dec}/L_{\rm drive} = 0.19 \pm 0.10$, when averaged over snapshots, Mach numbers, driving lengths, and the three LOSs.
 For LOS parallel to the magnetic field
the density structures are statistically smaller and the $L_{\rm dec}-L_{\rm drive}$ relation is tighter, with  $L_{\rm dec}/L_{\rm drive} = 0.112 \pm 0.024$.
We discuss our results in the context of using observations of chemical tracers to constrain the dominant turbulence driving scale in the ISM.

 \end{abstract}

\section{Introduction}
\label{sec: intro}

Understanding turbulence in galaxies is of central importance to a number of areas of astrophysical interest including star formation \citep[e.g.,][]{Krumholz2009,Ostriker10a,Burkhart2015}, cosmic ray acceleration and diffusion \cite{Schlickeiser02,LY14,Xu2016b}, and accretion disks around planets, stars and black holes \cite{Balbus91a,Hughes10a,Ross2017}. 
Compressible turbulence is ubiquitous throughout the interstellar medium (ISM) of galaxies from scales of at least tens of parsecs down to the sub-parsec scales \citep{Armstrong95, MckeeOstriker2007,Lazarian07rev,Chepurnov2010,krumreview2014,burkhartcollinslaz2015}.  
Turbulence in the ISM  may be driven by multiple energy injection sources on different scales \citep{Elmegreen2004,chepurnov15,Pingel2018}, from disk instabilities and supernova acting on the largest scales \citep{KrumholzBurkhart2018} to stellar winds and jets on sub-cloud scales \citep{Offner14a}.
Considering the wide range of scales galactic turbulence affects there has been significant effort to connect observed levels of turbulence with theoretical predictions and simulations \citep{GS95,Cho2003,Federrath2008,burkhart10,Correia2016,Herron2017}.  
% and dissipation scales. 
However, it is still unclear which driving mechanism  dominates the turbulent energy budget in the ISM \citep{krumreview2014} and on what scales the turbulence is dissipated \citep{Burkhart2015b}.

An important feature of a compressible turbulent cascade is that density fluctuations exhibit statistical correlations in relation to the driving scale \citep{Burkhart09a,Portillo2018}.
Numerical and analytic studies found that the correlation of density fluctuations decreases with increasing spatial separation ﻿ (\citealt{Kowal2007}, 
\citealt{Bialy2017}, hereafter \citetalias{Bialy2017}).
The characteristic scale over which the correlation decreases is the density decorrelation scale, $L_{\rm dec}$, and it is found to be of order of the driving scale, $L_{\rm drive}$,
\begin{equation}
\frac{L_{\rm dec}}{L_{\rm drive}} = \phi 
\label{eq:dec}
\end{equation}
with $\phi \approx 0.1 - 0.3$ (\citealt{VazquezSemadeni2001}, hereafter \citetalias{VazquezSemadeni2001}, \citealt{Fischera2004}, \citealt{Kowal2007}, \citetalias{Bialy2017}).
The exact value of $\phi$ depends on the method used to measure the decorrelation scale, for example, \citetalias{VazquezSemadeni2001} define $L_{\rm dec}$ as the point at which the autocorrelation function (ACF) falls to a fraction 0.1 of its initial value, whereas \citetalias{Bialy2017} derive $L_{\rm dec}$ using an analytic model that describes the correlation of the smoothed-density field as a function of the smoothing-length 
(see \S \ref{sec:theory} below; cf.~\citealt{Squire2017}).

Importantly, as discussed by \citetalias{Bialy2017} and \citet{Bialy2019}, $L_{\rm dec}$ may be constrained from 
observations of the column density probability density function (PDF) of various
atomic and molecular tracers (H, H$_2$, OH$^+$, OH$^+$, H$_2^+$, Ar$^+$).
This is because the chemical reactions in the ISM are sensitive to the gas density and its structure.
In particular, the absorption of  ultraviolet (UV) radiation by H$_2$ lines (i.e., H$_2$ self-shielding)
is very sensitive to the length-scales of density fluctuations.
In turn, other molecular species depend on the H$_2$ abundance, and are therefore also sensitive to $L_{\rm dec}$.
Given a robust relation between $L_{\rm dec}$ and $L_{\rm drive}$, 
 observations of chemical tracers may be used to constrain the turbulence driving scale (see Fig.~\ref{fig: diagram} and \S \ref{sub: implications to obs} below).

However, previous numerical studies have
obtained $L_{\rm dec}$ considering
only large driving scales, of order of the simulation box-size.
In this Letter, we use a large set of supersonic MHD simulations, 
driven on various scales: (1) large-scale $k_{\rm drive}=2.5$ driving (we denote the wavenumber $k \equiv 1/L_{\rm box}$), (2) intermediate scale $k_{\rm drive}=5$ driving, and (3) small scale $k_{\rm drive}=7$ driving,
and derive the $L_{\rm dec}-L_{\rm drive}$ relation as a function of $k_{\rm drive}$.
 For each driving-scale we further investigate the dependence of the density structures on the line-of-sight (LOS) orientation, parallel and perpendicular to the large scale magnetic field, 
and on the sonic Mach number.
%  generate MHD turbulence simulations with different driving  with all other initial conditions being similar and then apply the \citetalias{Bialy2017} method to measure the decorrelation scale. 

 This Letter is organized as follows: in \S \ref{sec:theory} we provide a theoretical overview for methods for deriving the decorrelation scale.
In \S \ref{sec: Numerics} we describe our numerical set up. 
In \S \ref{sec: Results} we present results for the density structures and the decorrelation scale, and their dependence on LOS orientation and driving scale. 
We discuss our results in \S \ref{sec: Discussion}, and conclude in \S \ref{sec: conclusions}.

\section{Theoretical Background}
\label{sec:theory}

The decorrelation scale,  $L_{\rm dec}$, is the characteristic scale over which density correlations decrease.
% For example, if the density correlations are determined solely by the turbulence driving process, then we might expect the ratio
% to be constant, independent of $L_{\rm drive}$.
% On the other hand, if other factors, such as magnetic fields, are important, we might expect $\phi$ to depend on $k_{\rm drive}$ and the LOS orientation.
We consider two definitions for $L_{\rm dec}$, (1) via the smoothed-density method (\citetalias{Bialy2017}), and (2) using the ACF (i.e., \citetalias[][]{VazquezSemadeni2001}).
% \begin{enumerate}
%     \item $L_{\rm dec}$ 
%     \item $L_{\rm dec}$ 
% \end{enumerate}{}
The smoothed-density method was developed for modeling the distribution of integrated column densities of chemical species and inferring properties of the 3D density field (\citetalias{Bialy2017}).
The idea is quite intuitive: 
if we smooth (average) the density over a scale $\ell$, and let $\ell$ vary from small to large, we expect the dispersion of the smoothed density to decrease with increasing $\ell$, as more density fluctuations are smoothed-out within the smoothing length

More quantitatively, given the field $x \equiv n/\langle n \rangle$ (i.e., normalized density), we define the smoothed-density
\begin{equation}
\label{eq: xl def}
    x_{\ell}(\ell) \equiv \frac{\int_{z}^{z+\ell} x \ \mathrm{d}z'}{\ell} \ ,
\end{equation}{}
where $\ell$ is the smoothing length and $z$ the line-of-sight (LOS) direction along which the density is smoothed.
If $x$ is a 3D field, then $x_{\ell}$ is also a 3D field but unlike $x$, $x_{\ell}$ also depends on $\ell$.
The distribution of $x_{\ell}$ is tightly related to that of the column density of slab of size $\ell$: $N= x_{\ell}\langle n \rangle \ell$.

Let $\sigma_x$ and $\sigma_{x\ell}(\ell)$ be the standard deviations (SDs) of $x$ and $x_{\ell}$.
To obtain an analytic description for $\sigma_{x\ell}(\ell)$, we assume
that the correlation may be described with a single parameter, $L_{\rm dec}$, such that when ${\ell} < L_{\rm dec}$ the density is correlated, while when $\ell \geq L_{\rm dec}$ the density is uncorrelated\footnote{This is obviously an approximation, as in a realistic density field that arises from turbulent cascade, the correlation does not fall abruptly, and it may vary with time and space. Nevertheless, as shown by \citetalias{Bialy2017}, this model provides a good approximation for the complicated density field found in the simulations.}.
The number of independent density cells along a LOS of length $\ell$ is
\begin{equation}
\label{eq: paz N}
\pazocal{N}(\ell) \approx \ell/L_{\rm dec} +1 \ \ \ ,
\end{equation}
and the $x_{\ell}$ distribution may be viewed as the sampling distribution of the mean (encountered in the error estimation of repeated measurements; \citealt{Barlow1989}).
The $x_{\ell}$ SD  obeys
\begin{equation}
\label{eq: sigma xL}
\frac{\sigma_{x_{\ell}}}{\sigma_x} = \frac{1}{\sqrt{\pazocal{N}(\ell)}} = \frac{1}{\sqrt{1+\ell/L_{\rm dec}}} \  \ \ .
% \simeq \frac{b \ms}{\sqrt{\pazocal{N}(\Delta z)}}  ,
\end{equation} 
In the limit $\ell/L_{\rm dec} \ll 1$, the smoothing length is smaller than a single density fluctuation, $\pazocal{N} \approx 1$, and $\sigma_{x_{\ell}} \rightarrow \sigma_x$. 
In the other extreme, when ${\ell}/L_{\rm dec} \gg 1$, $\pazocal{N} \gg 1$, many turbulent fluctuations are smoothed-over within $\ell$, and $\sigma_{x_{\ell}}/\sigma_{x}$ vanishes.
For more details and examples, see \S 4 in \citetalias{Bialy2017}.
See also \citet{Squire2017} for an alternative derivation of the density PDF as a function of scale.

Another way to define $L_{\rm dec}$ is from the ACF of the density field.
The ACF generally decreases with increasing lag, and we may define $L_{\rm dec}$ as the point at which the ACF falls below some fraction $\epsilon$ of its (initial) maximum value. 
This method depends on the somewhat arbitrary choice of $\epsilon$.
Interestingly, as we show below, the suggestion of \citetalias{VazquezSemadeni2001} to use $\epsilon=0.1$ yields $L_{\rm dec}$ values that are in very good agreement with those obtained via our smoothed-density method  (\S \ref{sub: results Ldec-Ldrive}).

% We define the ratio
% \begin{equation}
%     \phi \equiv \frac{L_{\rm dec}}{L_{\rm drive}} \ ,
% \end{equation}{}
% where $L_{\rm drive}$ is the turbulence driving scale.
% In \S \ref{sec: Results} we explore the $\phi$ value as calculated in a set of MHD simulations, and whether its value is a constant or if it varies with
% the driving scale and line-of-sight (LOS) orientation.

\begin{figure*}[t]
	\centering
	\includegraphics[width=1\textwidth]{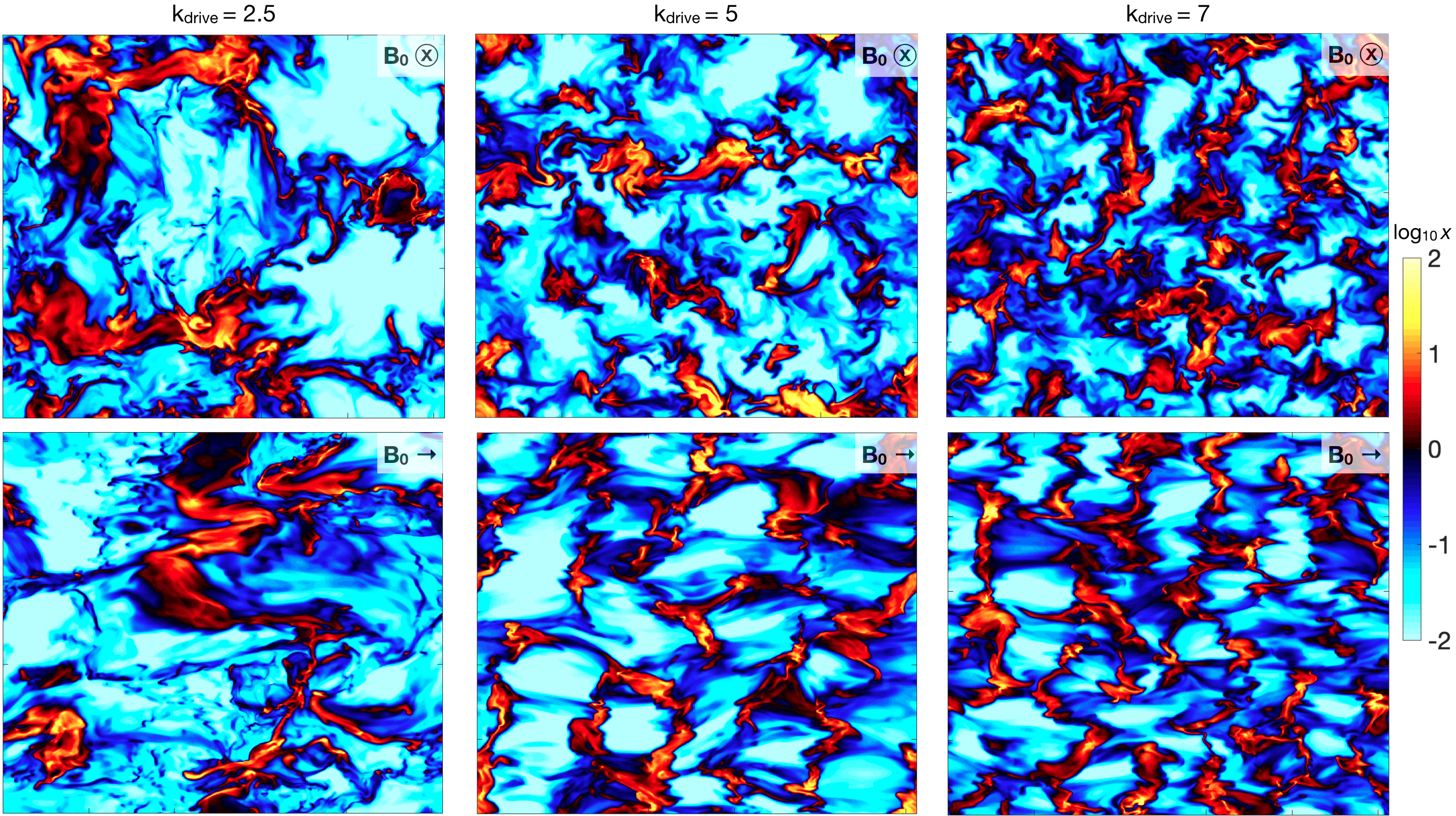} 
	\caption{
Density cuts through the $\mathcal{M}_s=4.5$, $k_{\rm drive}=2.5$ (left), $k_{\rm drive}=5$ (middle) and $k_{\rm drive}=7$ (right) simulations.
In the top panels ${\bf B_0}$ is directed into the plane and in the bottom from left to right.
The colorscale corresponds to $\log_{10} x \equiv \log_{10} n/\langle n \rangle$.
}    %/Turb//plots_for_paper/profiles_and_N1_tot_vs_
\label{fig: simulations}
\end{figure*}

\section{Numerical Method}
\label{sec: Numerics}

\subsection{MHD simulations}
\label{sub: MHD sims}
We run 3D numerical simulations of
isothermal compressible MHD turbulence.
 The code and setup is similar to that of a number of past works (\citealt{Kowal2007}, \citealt{Burkhart09a}, \citetalias{Bialy2017}).
We refer to these works for the
details of the numerical set-up and here provide a short
overview.  
% We use the isothermal MHD code detailed in .
The code is a third-order accurate
ENO scheme which solves the ideal MHD equations
in a periodic box with purely solenoidal driving \citep{Cho03}.
The magnetic field consists of the uniform background
field and a turbulent field, i.e: ${\bf B} = {\bf B_0} + {\bf b}$ with the magnetic field initialized along a single preferred direction.
We run two sets of simulations, with a sonic Mach number $\mathcal{M}_s = 4.5$ and $\mathcal{M}_s = 7$.
While previous studies used driving on large scales, with $k_{\rm drive}=2-2.5$, here, 
for each Mach number, we run several simulations, each of which driven on a different driving scale, of $k_{\rm drive}=2.5, 5$, and $7$.
We also ran simulations of $k_{\rm drive}=10$ but for this high $k_{\rm drive}$ the results do not robustly converge and thus we do not discuss this simulation further.
The  Alfv\'enic Mach number in all the simulations is 
$\mathcal{M}_A= 0.7$.
%The simulations are part of the Catalog for Astrophysical Turbulence Simulations (CATS\footnote{See \href{www.mhdturbulence.com}{www.mhdturbulence.com}.}, Burkhart et al. 2019)
To test numerical convergence we 
also run simulations with various resolutions. For the $\mathcal{M}_s=4.5$ simulations we run, $N_{\rm res}=256^3, 512^3$ and $1024^3$ resolution elements. For the $\mathcal{M}_s=7$ simulation we only run  $1024^3$ resolution. 
% As we discuss in \S \ref{sub: calc Ldec}, in our analysis of $L_{\rm dec}$, we consider 5 time snapshots for each simulations, to evaluate  statistical errors.
% In total, we analyze 
% {\color{magenta} $3 \times 3 \times 5 = 45$ density fields.
% }
% For each density field we compute the function $\sigma_{x\ell}(\ell)$ and $L_{\rm dec}$
% along three LOS orientations, as discussed in \S \ref{sub: calc Ldec} below.

%  and 
% set the

% $M_s=4.5$ and $M_A=0.7$. 

\subsection{Calculating $L_{\rm dec}$ for the MHD boxes}
\label{sub: calc Ldec}
For each simulation, characterized by 
$(\mathcal{M}_s, k_{\rm drive},N_{\rm res})$
we calculate $L_{\rm dec}$ as follows:
% \begin{enumerate}[label=(\Alph*)]
    % \item We calculate $L_{\rm dec}$:
        \begin{enumerate}[label=(\arabic*)]
            \item calculate $\sigma_x$ for that simulation.
            \item calculate $\sigma_{x\ell}(\ell)$: 
            % \begin{enumerate}[label=(\roman*)]
                % \item
                We choose $\ell$, and integration orientation (hereafter denoted by line-of-sight, LOS). 
                We pick $5 \times 10^5$ random locations (cells) in the simulation and for each location we compute the smoothed density $x_{\ell}$ using Eq.~(\ref{eq: xl def}) (we use periodic boundaries). This gives the $x_{\ell}$ distribution at scale $\ell$.
                % \item 
                We repeat this
                % step (i) 
                for $\ell$ values ranging from 0 to 1 (we adopt units normalized to the box length)
                and calculate $\sigma_{x\ell}$ as a function of $\ell$.
            % \end{enumerate}
            \item 
             Fit Eq.~(\ref{eq: sigma xL}) to the numerical data, $\sigma_{x\ell}(\ell)/\sigma_{x}$ as a function of $\ell$, with $L_{\rm dec}$ being the best-fitting parameter that minimizes $\chi^2$.
        \end{enumerate}
    % \item 
    We follow the procedure above for three LOS orientations, 1 parallel and 2 perpendicular to ${\bf B_0}$.
    For each simulation and LOS orientation, we repeat the steps above 5 times for 5 time snapshots and  adopt the average $L_{\rm dec}$ as the value of the decorrelation scale.
    For the error we sum in quadrature the error from the $\chi^2$ fitting (step 3), and the SD $L_{\rm dec}$ over the 5 time snapshots. 
    % \item 
    In conclusion we obtain $L_{\rm dec}\pm \Delta L_{\rm dec}$ for different $\mathcal{M}_s$, $k_{\rm drive}$, $N_{\rm res}$, and LOS orientation.
    % \end{enumerate}

% \begin{figure*}[t]
% 	\centering
% 	\includegraphics[width=\textwidth]{L_dec} 
% 	\caption{
% 	The deccorelation to driving scale ratio $L_{\rm dec}/L_{\rm drive}$, as a function of the driving number, $k_{\rm drive}$.
% For each $k_{\rm drive}$, the blue and black points are for LOS perpendicular to the $B$ field, and the red points are for LOS along the field.
% Each point is an average over the different time snapshots, and the error bars correspond to a $1 \sigma$ spread about the average.
% }    %/Turb//plots_for_paper/profiles_and_N1_tot_vs_
% \end{figure*}

\begin{figure}[t]
	\centering
	\includegraphics[width=0.5\textwidth]{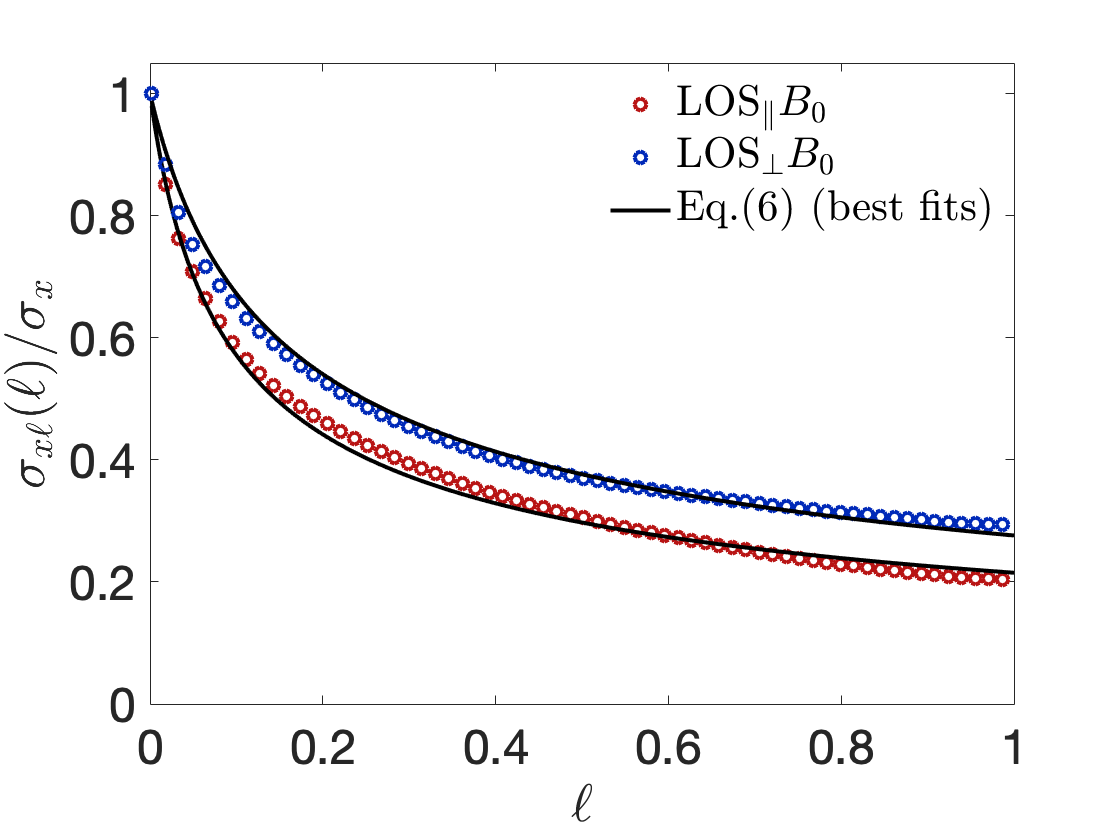} 
	\caption{
The smoothed-density SD normalized to the density SD, 
as a function of smoothing length, $\ell$ (in units of box-length), for the $\mathcal{M}_s=4.5$, $k_{\rm drive}=2.5$, $N_{\rm res}=1024^3$ simulation.
The red and blue points correspond to LOS$_{\parallel }{\bf B_0}$ and $_{\bot }{\bf B_0}$, and the black curves are fits to Eq.~(\ref{eq: sigma xL}), yielding $L_{\rm dec}=(4.85,8.25) \times 10^{-2}$, respectively.
}    %/Turb//plots_for_paper/profiles_and_N1_tot_vs_
\label{fig: fit}
\end{figure}

\begin{figure*}[t]
	\centering
	\includegraphics[width=1\textwidth]{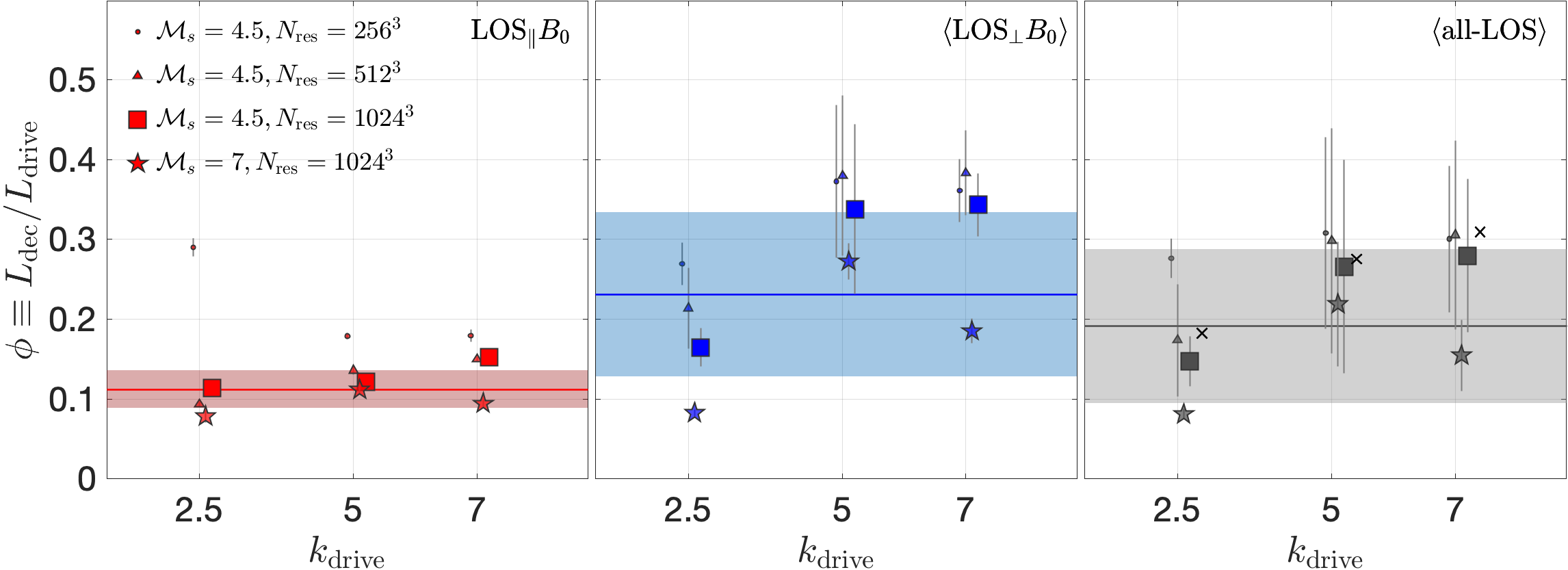} 
	\caption{
	The decorrelation-scale to driving-scale ratio, $\phi \equiv L_{\rm dec}/L_{\rm drive}$ as a function of driving wavenumber, $k_{\rm drive}\equiv1/L_{\rm drive}$ (in box-length units), as calculated for our set of MHD simulations using the smoothed density method.
	The three panels correspond to the LOS$_{\parallel }{\bf B_0}$ (left), an average over the LOS$_{\bot }{\bf B_0}$ (middle), and all-LOS average (right).
	The error bars combine fitting uncertainly and the dispersion over simulation time snapshots and the LOS averaging. 
	The cross symbols in the right panel show the $\phi$ computed using the auto-correlation function method, for the  $\mathcal{M}_s=4.5$ simulations.
    In each panel, the horizontal line is the mean $\phi$ over $k_{\rm drive}$ and $\mathcal{M}_s$, and the shaded strip is the $\pm 1 \sigma$ uncertainty range.
}    
\label{fig: Ldec}
\end{figure*}

\section{Results}
\label{sec: Results}

In this section we present results for the density structures in the turbulent boxes, and particularly the dependence of the density decorrelation scale, $L_{\rm dec}$, on driving scale, 
LOS orientation, and resolution.

\subsection{Density Slices}
\label{sub: results den slices}
We start with some visual examples of the data.
In Fig.~\ref{fig: simulations} we show density slices for the 
$\mathcal{M}_s=4.5, N_{\rm res}=1024^3$, 
$k_{\rm drive}=2.5$ (left), 5 (middle), and 7 (right) simulations.
% through 3 three simulations in our set, with $k_{\rm drive}=2.5$, 5, and 7 (all with $N_{\rm res}=1024$).
% The color axis shows $\log_{10} x \equiv \log_{10} (n/\langle n \rangle)$ ranging from -2 (cyan) to +2 (yellow) (see colorbar).
Comparing the panels left-to-right, it is evident that density structures are typically smaller as the driving scale decreases.
This is expected as the density fluctuations develop as a result of the driving process.
In the upper panels, we see that density structures are relatively isotropic (compared to the lower panels).
This is because in these panels ${\bf B_0}$ is directed into the plane and thus there is no preferred direction. Thus the structure is more reminiscent of pure hydro turbulence.
On the other hand in the lower panel, where ${\bf B_0}$ is in the plane, the density structures are not isotropic and tend to have their shorter dimension along ${\bf B_0}$. 
This behavior makes sense physically as the gas may stream more freely in directions along the magnetic field and thus gas compressions are more efficient.
As we show in \S \ref{sub: results Ldec-Ldrive}, our calculated decorrelation scale as a function of $k_{\rm drive}$ and LOS orientation captures these trends in a quantitative manner.

\subsection{The standard deviation of the smoothed and non-smoothed density}
\label{results: sigma_xl}
In Fig.~\ref{fig: fit} we show an example of the calculated 
$\sigma_{x\ell}/\sigma_x$ as a function of the smoothing length $\ell$, for the 
$\mathcal{M}_s=4.5$, $k_{\rm drive}=2.5$, $N_{\rm res}=1024^3$ simulation.
The red points correspond to the LOS$_{\parallel }{\bf B_0}$ and the blue points to a LOS$_{\bot }{\bf B_0}$.
The black curves are the best $\chi^2$ fits to Eq.~(\ref{eq: sigma xL}) which
yield $L_{\rm dec}$ for these LOS orientations.
As expected, $\sigma_{x\ell}/\sigma_x$ approaches unity in limit $\ell \rightarrow 0$ as smoothing becomes ineffective.
As $\ell$ increases, more of the density structures are averaged-out and $\sigma_{x\ell}$ decreases.
$\sigma_{x\ell}$ falls faster for the LOS$_{\parallel }{\bf B_0}$ than that of the LOS$_{\bot }{\bf B_0}$
as the density structures are non-isotropic and are typically shorter along the direction of ${\bf B_0}$ (see \S \ref{sub: results den slices} and Fig.~\ref{fig: simulations}).
The corresponding $L_{\rm dec}$ is thus smaller for LOS$_{\parallel }{\bf B_0}$, with $L_{\rm dec}=4.9 \times 10^{-2}$ and $8.3 \times 10^{-2}$ for LOS$_{\parallel }{\bf B_0}$ and LOS$_{\bot }{\bf B_0}$, respectively.
As we show in \S \ref{sub: results Ldec-Ldrive}, this difference remains also after time averaging and is seen in all simulations, from small to large $k_{\rm drive}$ and for both considered Mach numbers.

 \subsection{The decorrelation-scale driving-scale relation}
 \label{sub: results Ldec-Ldrive}
In Fig.~\ref{fig: Ldec} we show the ratio $\phi \equiv L_{\rm dec}/L_{\rm drive}$ as a function $k_{\rm drive}$ for various resolutions 
and sonic Mach numbers
(different symbols), and LOS orientation: the LOS$_{\parallel }{\bf B_0}$ (left panel), the average over the two LOS$_{\bot }{\bf B_0}$ (middle) and an average over all three LOSs (right).
As discussed in \S \ref{sub: calc Ldec} each $L_{\rm dec}$ is also an average over 5 time snapshots and the error bars 
are $\pm 1 \sigma$ uncertainty ranges, corresponding to quadrature sums of the fitting process error and the SD over the time snapshots. For the left and right panels, the errors are larger as they also include the dispersion over the averaged LOS.

If correlations in the density field are imposed by the driving scale, we expect $\phi \equiv L_{\rm dec}/L_{\rm drive}$ to be constant in respect to $k_{\rm drive}$.
Starting from the left panel of  Fig.~\ref{fig: Ldec} we see that in the case of the LOS$_{\parallel }{\bf B_0}$, $\phi$ indeed remains nearly constant across $k_{\rm drive}$ and $\mathcal{M}_s$.
The average $\phi$ over $k_{\rm drive}$ and $\mathcal{M}_s$ (at $N_{\rm res}=1024^3$) is
\begin{equation}
    \langle \phi \rangle_{{\parallel }{\bf B_0}} = 0.112 \pm 0.024 \ .
\end{equation}
The error is the $\pm 1\sigma$ uncertainty range, 
encompassing the dispersion (the SD) across the driving scale, Mach number, simulation time snapshot, summed in quadrature with the fitting procedure error.
% the dispersion (the SD) of all the points across the $k_{\rm drive}$ and $\mathcal{M}_s$ range, and the error bars on the points (i.e., the error bars in Fig.~\ref{fig: Ldec}) which arises from time averaging and fitting uncertainty.
The mean and the $\pm 1 \sigma$ uncertainty range are shown as the red horizontal line and shaded strip in Fig.~\ref{fig: Ldec}.

For the LOS$_{\bot }{\bf B_0}$ (middle panel) the $\phi$ values are higher, with
\begin{equation}
    \langle \phi \rangle_{{\bot }{\bf B_0}} = 0.23 \pm 0.11 \ ,
\end{equation}
 and have a larger uncertainty range.
The larger $L_{\rm dec, \bf \bot B_0}$, compared to $L_{{\rm dec}, {\parallel }{\bf B_0}}$, may be seen by-eye in the density slices
presented in Fig.~\ref{fig: simulations}, and may be explained by the fact that gas compression is limited in the direction perpendicular to the magnetic field (see \S \ref{sub: results den slices}).
The SD deviation is also much larger in the $_{{\bot }} {\bf B_0}$ case.
Furthermore, $\phi$ shows an increasing trend with increasing $k_{\rm drive}$.
However, numerical convergence is not optimal for these LOS (as evident by comparing the various resolution markers), and the number of points across $k_{\rm drive}$ is limited.
A $\phi$ that increases with $k_{\rm drive}$ may be expected if the large scale ${B_0}$ field induces correlations on large scales proportional to the simulation box length rather than the driving scale. 

Finally, in the right panel we show the average $L_{\rm dec}$ over all LOS (with appropriate weights: 2/3 for the LOS$_{\bot }{\bf B_0}$ and 1/3 for the LOS$_{\parallel }{\bf B_0}$).
We obtain
\begin{equation}
    \langle \phi \rangle_{\rm all \ LOS} = 0.19 \pm 0.10 \ ,
\end{equation}

We also calculated the decorrelation lengths from the ACF, by finding the point at which the ACF falls to a fraction $\epsilon=0.1$ of its maximal value.
This measure was suggested by \citetalias{VazquezSemadeni2001} in their study of column density PDFs (which are tightly related to the $x\ell$ distribution).
In Fig.~\ref{fig: Ldec} we compare $\phi$ as computed using our smoothed-density method and via the the ACF, for the $\mathcal{M}_s=4.5$ simulations.
Interestingly, the decorrelation length obtained from the ACF agrees well with that obtained with our smoothed-density method.
However, while the ACF method depends on the arbitrary choice of $\epsilon$, our method does not require any tuning as it relies on an analytic model that describes the dependence of $\sigma_{x\ell}(\ell)/\sigma_x$ on $L_{\rm dec}$ (see \S \ref{sec:theory}).

\begin{figure*}[t]
	\centering
	\includegraphics[width=0.85\textwidth]{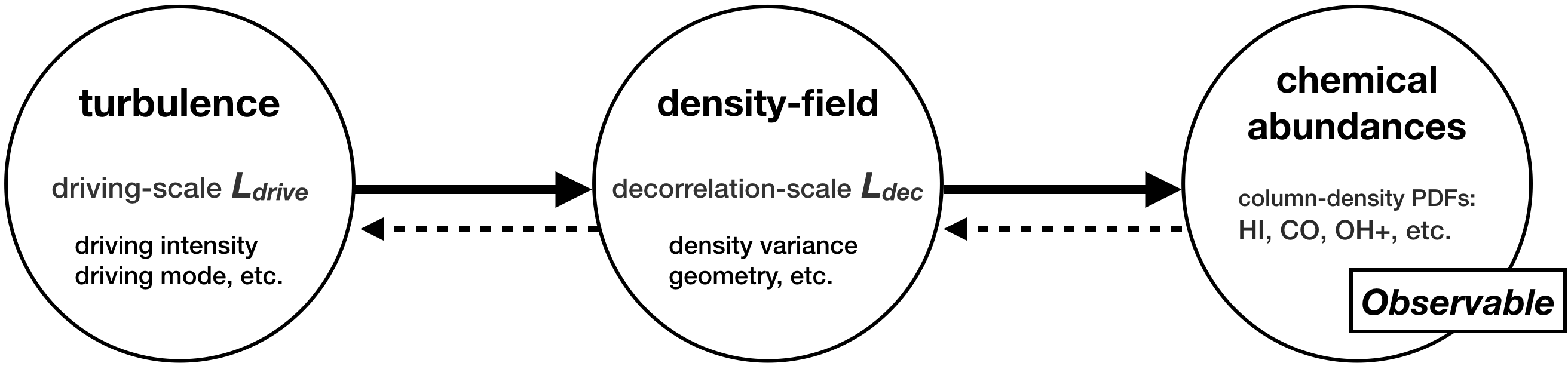} 
	\caption{
    Schematic diagram demonstrating how turbulence determines the density field which in turns affect the PDF of chemical abundances (solid arrows).
    Thus observations of column density PDFs may be used to constrain the density field and turbulence properties (dashed arrows).
    Therefore it is important to quantify the connections between properties of turbulence driving, the density field, and chemical structure.
        }    
\label{fig: diagram}
\end{figure*}

\section{Discussion}
\label{sec: Discussion}

In this Letter we have explored the density structures that arise in a supersonic magnetized driven turbulence box simulations.
In particular, we focused on quantifying the relation between the decorrelation scale ($L_{\rm dec}$) and the turbulence driving scale ($L_{\rm drive}$) as well as on the orientation relative to the large scale magnetic field.
We find that the $L_{\rm dec}-L_{\rm drive}$ relation may be approximated by a constant ratio, of $\phi \equiv L_{\rm dec}/L_{\rm drive} = 0.19$, on average.
If only the LOS parallel to ${\bf B_0}$ is considered, the decorrelation scale is smaller, with $\phi = 0.11$, and the relation is tighter.

\subsection{Implications to Observations}
\label{sub: implications to obs}
In a broader context, the $L_{\rm dec}-L_{\rm drive}$ relation 
is a key component in the quest to constrain turbulence properties from observations.
This is depicted in Fig.~\ref{fig: diagram}.
The diagram shows that the turbulence, the density structure, and the chemical structure of interstellar gas are connected:
\begin{enumerate}[label=(\Alph*)]
    \item Turbulence (when supersonic) produces strong density fluctuations in the gas such that the properties of the density field depend on the turbulence properties
    \item The density structure, in turn, controls the abundances of various chemical species since the rates of chemical reactions are sensitive to gas density.
\end{enumerate}{}
Thus, we may potentially use observations of chemical abundances 
to constrain the density field and turbulence properties.
For this we need to quantify the (A) and (B) connections in Figure 4 (i.e., the solid arrows).
In previous work we investigated connection (B), focusing on how $L_{\rm dec}$ and the sonic Mach number control the abundances of HI (\citetalias{Bialy2017}), and the molecular ions OH$^+$, H$_2$O$^+$, and ArH$^+$ \citep[][]{Bialy2019}.
In this paper we focused on connection (A) in Figure 4, and established the link between the decorrelation-scale, $L_{\rm dec}$, and the driving-scale, $L_{\rm drive}$, via a set of MHD simulations driven on varying scales. 
More generally, 
turbulence driving is also described by other parameters, such as the velocity dispersion at the driving scale, and the ratio of solenoidal versus compressional modes, which also affect the density field.

\subsection{Limitations and Future Work}
\label{sub: limitations}
In this study we have analyzed 3D MHD driven box simulations with an isothermal equation of state (see \S \ref{sub: MHD sims}).
In the realistic ISM, the density field is affected by active cooling and heating processes, which render the equation of state non-isothermal, and leading to the formation of a multiphase medium composed of cold-dense and warm-diffuse gas \citep{Field69a, Wolfire2003, Bialy2019phases}, although the phase separation vanishes when turbulence is sufficiently strong \citep{Gazol2013, Kritsuk2017}.
In a future study, it would be interesting to investigate the structure of the density field (i.e., the $L_{\rm dec}-L_{\rm drive}$ relation) in a non-isothermal medium.
Other important generalizations is a self-gravitating medium, and the inclusion of feedback (i.e., supernova feedback), which can drive turbulence and provide gas heating.
In this study we deliberately used the setup of isothermal, non-gravitational, Fourier-driven simulations, as they constitute a clean numerical experiment that are useful for deriving and understanding the basic form of the $L_{\rm dec}-L_{\rm drive}$ relation.

\section{Conclusion}
\label{sec: conclusions}
We found that the decorrelation-scale of the density field, $L_{\rm dec}$, is related to the turbulence driving scale, following an approximately constant ratio, $L_{\rm dec}/L_{\rm drive} \approx \phi$ where $\phi = 0.112 \pm 0.024$ for density fluctuations along the large scale ${\bf B_0}$ field, and with 
$\phi = 0.23 \pm 0.11$ for fluctuations perpendicular to ${\bf B_0}$.
On average over all directions, $\phi = 0.19 \pm 0.10$.
The decorrelation scale calculated with our smoothed density method is in good agreement with that obtained from the autocorrelation function.
The $L_{\rm dec}-L_{\rm drive}$ relation is a key step for constraining the turbulence driving scale from observations of column density PDFs.
This may shed light on the relative importance of various turbulence stirring mechanisms in the Galaxy.

% \begin{enumerate}
%     \item Implications to observations: The Width of a column density PDF depends on $L_{\rm drive}$ or $L_{\rm dec}$
%     \item PPV.
%     \item Models: H-H2
%     \item The possibility to measure the driving scale from PDFs
%      \item  There can be several drivings in the real ISM
% \end{enumerate}

\acknowledgements
We are grateful for valuable discussions with Amiel Sternberg.
We thank the referee for useful suggestions that improved this paper.
B.B. acknowledges the generous support of the Simons Foundation Flatiron Institute Center for Computational Astrophysics (CCA). S.B acknowledges support from the Harvard-Smithsonian Institute for Theory and Computation (ITC) and visitor support from the CCA.  

%Blakesley: I should also thank the CCA which hosted me several times where we worked on thiis project. 
% is there a right way for such an acknowledgement?....I added, take a look.

\end{document}